\documentclass[fleqn,10pt]{wlscirep}
\title{Stochastic Spin-Orbit Torque Devices as Elements for Bayesian Inference}

\author[1]{Yong Shim}
\author[1]{Shuhan Chen}
\author[1,*]{Abhronil Sengupta}
\author[1]{Kaushik Roy}
\affil[1]{School of Electrical \& Computer Engineering, Purdue University, West Lafayette, Indiana 47907, USA}

\affil[*]{asengup@purdue.edu}

\begin{abstract} 
Probabilistic inference from real-time input data is becoming increasingly popular and may be one of the potential pathways at enabling cognitive intelligence. As a matter of fact, preliminary research has revealed that stochastic functionalities also underlie the spiking behavior of neurons in cortical microcircuits of the human brain. In tune with such observations, neuromorphic and other unconventional computing platforms have recently started adopting the usage of computational units that generate outputs probabilistically, depending on the magnitude of the input stimulus. In this work, we experimentally demonstrate a spintronic device that offers a direct mapping to the functionality of such a controllable stochastic switching element. We show that the probabilistic switching of Ta/CoFeB/MgO heterostructures in presence of spin-orbit torque and thermal noise can be harnessed to enable probabilistic inference in a plethora of unconventional computing scenarios. This work can potentially pave the way for hardware that directly mimics the computational units of Bayesian inference. 
\end{abstract} 

\begin{document}

\flushbottom
\maketitle
%
%
\thispagestyle{empty}


\section*{Introduction}

Spin-orbit torque generated by an underlying heavy metal has recently emerged as an energy-efficient mechanism for magnetization reversal \cite{liu2012spin,miron2011perpendicular,liu2012current} and domain wall motion \cite{ryu2013chiral,ryu2014chiral,emori2013current,emori2014spin}. Nanomagnet switching due to input charge current flowing through the heavy-metal (HM) underlayer is mainly attributed to spin-Hall effect (SHE) \cite{hirsch1999spin}, wherein, a transverse spin current is injected in the nanomagnet lying on top. While magnets with in-plane anisotropy can be switched directly by spin-orbit torque (SOT), perpendicular magnets require an external magnetic field for deterministic switching. Recent proposals have also explored deterministic switching in perpendicular magnets without the assistance of any external field \cite{yu2014switching,you2015switching,torrejon2015current}. While such spin-orbit torque induced magnetization switching has been extensively studied in the deterministic regime, it is intrinsically probabilistic due to the inherent time-varying thermal noise involved in the magnetization dynamics. 

This work firstly attempts to experimentally validate prior theoretical proposals for utilizing spin-orbit torque switching nanomagnets as biased random number generators where the bias can be tuned using the magnitude of the input stimulus by operating the magnets in the stochastic regime \cite{sengupta2016magnetic,srinivasan2016magnetic,sengupta2016probabilistic,sutton2016intrinsic,shim2016ising}. 
Based on the experiment, we extend the concept to a three-terminal device structure that can be easily interfaced with CMOS peripherals for enabling different genres of unconventional computing scenarios. The device-circuit configuration based on the stochastic spin-device and CMOS interface circuits forms a core hardware primitive and used for versatile applications ranging from neuromorphic \cite{sengupta2016magnetic,srinivasan2016magnetic,sengupta2016probabilistic} to combinatorial optimization \cite{sutton2016intrinsic,shim2016ising}. As a second contribution, we propose a Bayesian inference engine by exploiting the controllable stochastic switching of the nanomagnets. Each nanomagnet along with peripheral CMOS circuits form a key element, a variable, of the Bayesian Network (BN) and generates Poisson spike pulse train. The probabilistic information is transferred through the interconnected variables by following pulse based arithmetic \cite{murray1988pulse}. The efficiency of such spintronic-enabled probabilistic Bayesian networks stems from the direct mapping of the key stochastic computing element to the underlying stochastic device physics of the spin devices.

\section*{Device fabrication and spin-orbit torque (SOT) driven stochastic switching}

The probabilistic switching was characterized in a $1.2 \mu m$ wide Ta (10 nm)/CoFeB(1.3 nm)/MgO (1.5nm) Hall-cross structure in presence of a $100$ $Oe$ in-plane external magnetic field. The in-plane magnetic field is required to achieve switching of the PMA (perpendicular magnetic anisotropy) free layer in presence of in-plane polarized spins generated by current flowing through the heavy metal underlayer. 

\begin{figure*}
\centering
\includegraphics[width=7.0in]{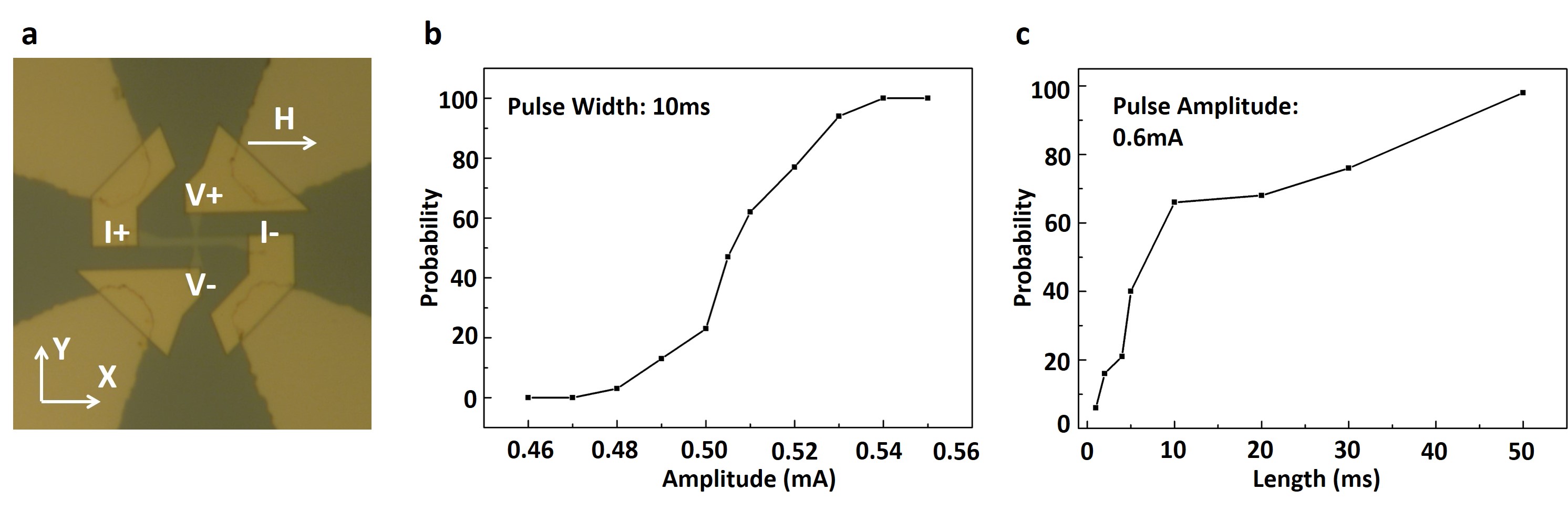}
\caption{(a) Nanoscale Hall-bar structure consisting of Ta (10 nm)/CoFeB (1.3 nm)/MgO (1.5 nm)/Ta (5 nm) (from bottom to top) material stack. Input current flows between terminal $I+$ and $I-$ while the magnetization state is detected by change in the anomalous Hall-effect resistance measured between terminals $V+$ and $V-$. The device is subjected to an in-plane external magnetic field of magnitude $100$ $Oe$. (b) Variation of the magnet switching probability with variation in magnitude (amplitude) of the input current pulse. The pulse width was fixed at 10ms. The switching probability is $0\%$ at 0.47mA and $100\%$ at 0.54mA. (c) Variation of the magnet switching probability characteristics with variation in the duration of the input current pulse. The pulse amplitude is fixed at 0.6mA. The switching probability is $6\%$ at 1ms while it attains a value of $98\%$ at 50ms. Note that the switching probability characteristics depicted in (b) and (c) are for two different samples and similar behavior was repeatedly observed over other fabricated devices.}
\label{fig1}
\end{figure*}

Fig. \ref{fig1}(a) depicts the Hall-bar structure. Input current flows between the terminals $I+$ and $I-$ while the magnetization state is determined by the anomalous Hall effect resistance detected between terminals $V+$ and $V-$. Initially the magnet is reset by passing a sufficient magnitude of current between $I+$ and $I-$ terminals in the negative x direction. Subsequently, a current is passed in the positive x direction and the final state of the magnet is determined. The magnitude of the current is varied and over 50 measurements are taken per current magnitude to determine the probability of switching of the magnetic stack. The experiment was repeated over multiple devices and consistent probability switching characteristics were obtained. Fig. \ref{fig1}(b) represents the variation of switching probability of the magnet with variation in the magnitude of the input current pulse (with the pulse width being fixed at 10 ms) while (c) depicts that the switching probability can be tuned by varying the pulse width as well (with pulse amplitude being fixed at 0.6mA). 
Prior proposals have exploited such non-linear switching probability characteristics for neuromorphic \cite{sengupta2016magnetic,srinivasan2016magnetic,sengupta2016probabilistic} as well as Ising computing \cite{sutton2016intrinsic,shim2016ising}. Additionally pulse width duration serves to control the rate of change of switching probability with variation in the input current magnitude which, in turn, impacts the performance of such computing systems in presence of noise and process variations in the spin-devices and CMOS peripherals \cite{sengupta2016probabilistic}. 

\subsection*{Extension to three-terminal device and possible applications}

\begin{figure*}
\centering
\includegraphics[width=7.0in]{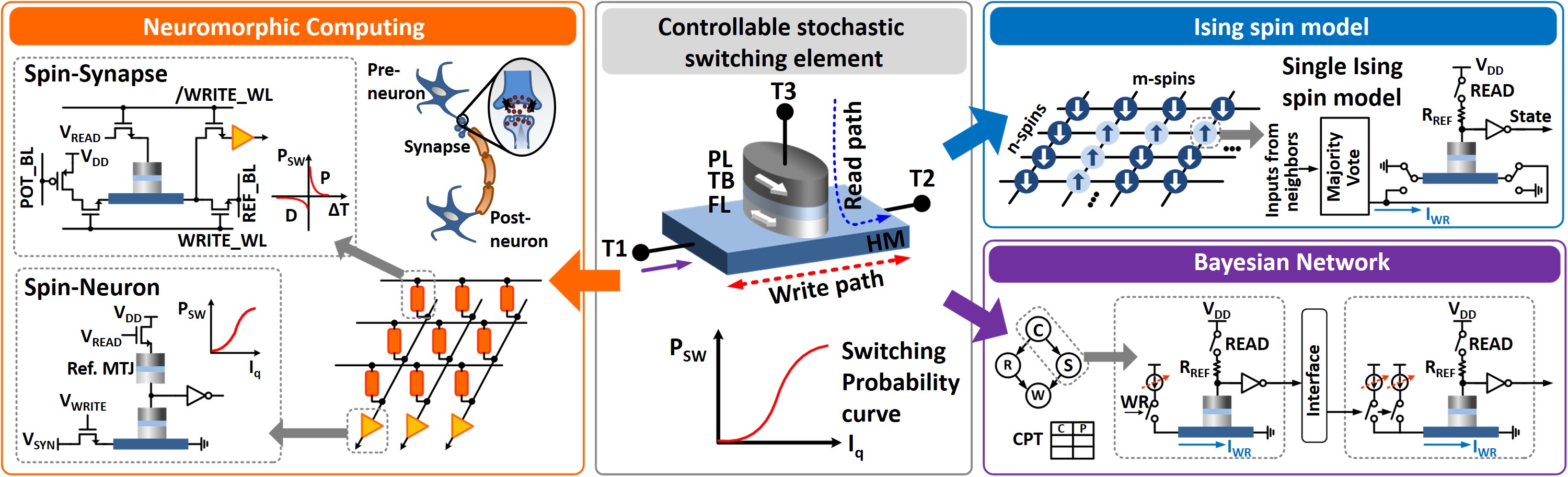}
\caption{A three terminal device structure consists of Magnetic Tunnel Junction (MTJ) on top of the Heavy Metal (HM) underlayer.(Center) The stochastic switching of the device in the presence of thermal noise can be exploited to various applications such as neuromorphic computing (Left), Ising spin model (Right top), and Bayesian Network (Right bottom). }
\label{fig3}
\end{figure*}

The experiment in the previous section provides proof-of-concept for a large number of theoretical proposals that have exploited probabilistic SOT-driven magnetization dynamics in a three-terminal device structure shown in Fig. \ref{fig3} (center). The three terminal device structure, shown in Fig. \ref{fig3}, can be designed by fabricating a Magnetic Tunnel Junction (MTJ) stack on top of the ferromagnet-heavy metal layers. 
Note that, the conventional MTJ device consists of two ferromagnetic layers, which are separated by a Tunneling Barrier (TB) between them. The top layer is called the Pinned Layer (PL) and the bottom layer is called the Free Layer (FL). The magnetization of the FL can be manipulated by an input spin current injected from the heavy metal underlayer while the magnetization of the PL is fixed in a particular direction. When the magnetization of the two ferromagentic layers is located in the same direction, the resistance across the tunneling junction has a smaller resistance (R$_{P}$) compared to the opposite case (R$_{AP}$). While the write current flows between terminals T1 and T2 and probabilistically switches the magnet (probability determined by current magnitude), the read path between terminals T3 and T2 determines the final state of the magnet after the switching process. 

Note that the only difference between the fabricated samples and the device structure (shown in Fig. \ref{fig3}) is the read-out operation. While the magnet switching dynamics is similar to the fabricated samples, through the injection of spin current, the read-out mechanism is through a MTJ structure lying on top of the heavy metal. The tunnel junction exhibits a much larger resistance variation (typically 2-3 times) corresponding to the two stable magnetization states that, in turn, leads to compatibility with peripheral CMOS technology. The subsequent text and applications discussed in this article are based on the device measurements shown in Fig. \ref{fig1}. Consequently, all conclusions presented in this article are based on experimentally measured stochastic switching characteristics of the magnet. However, for performing the CMOS circuit-level simulations we consider that the device resistance that can be sensed is similar to values obtained from standard MTJ stacks compatible with CMOS technology. Such devices can be scaled down to dimensions exhibiting barrier height of the order of $\sim 10-20 k_{B}T$. Further scaling can potentially result in device state update during ``read” operation as the device becomes increasingly sensitive to the input bias current. 

The controllable stochastic switching element can be potentially used in various applications as shown in Fig. \ref{fig3}.
For instance, neuromorphic applications inspired by brain functionalities consist of a set of pre-neurons transmitting information to a set of post-neurons through synapses. Such a computing framework can be mapped to a crossbar array of stochastic switching elements (serving as synapses) driving stochastic post-neuronal devices. The device can be interfaced with peripheral transistors to implement stochastic Spike-Timing Dependent Potentiation (P) and Depression (D) learning rules which dictate the variation of synaptic switching probability with spike timing $\Delta T$. Similarly, stochastic neural functionality can be implemented by interfacing the neuronal device with a Reference MTJ. For details, please refer to Refs. \cite{sengupta2016magnetic,srinivasan2016magnetic,sengupta2016probabilistic}. 
Stochastic switching property of the device can also provide a natural annealing property in Ising computing systems for solving optimization tasks by assisting the system to move out of a local minima \cite{sutton2016intrinsic,shim2016ising}. 
Here we explore Bayesian inference networks and utilize the probabilistic switching characteristics in response to pulse current magnitude of the sample, depicted in Fig. \ref{fig1}(b), as the core enabling element for probabilistic inference.

\subsection*{Micromagnetic simulation for device modeling}
\begin{figure*}
\centering
\includegraphics[width=5in]{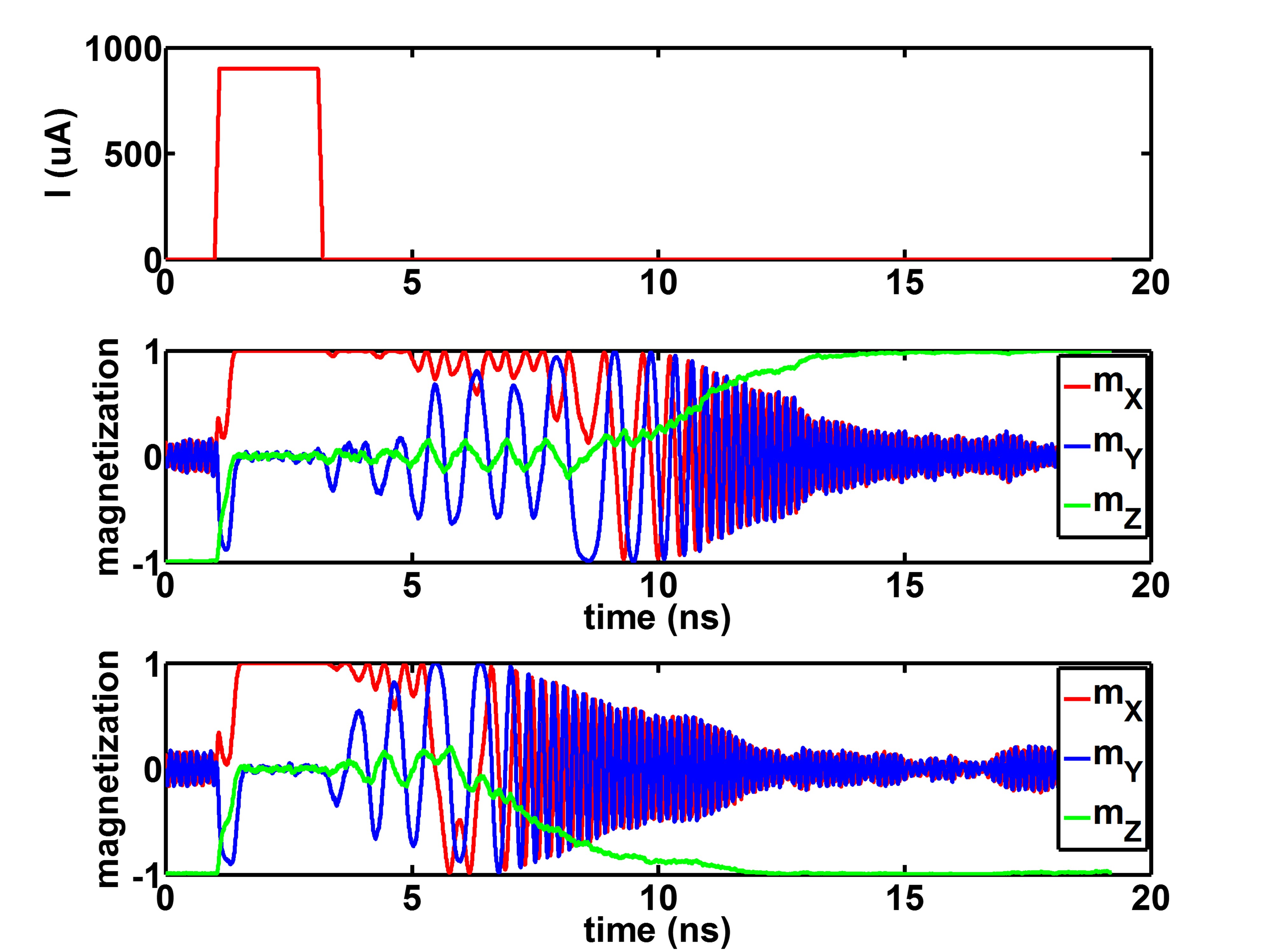}
\caption{Two independent simulations of stochastic Landau-Lifshitz-Gilbert (LLG) equation with thermal noise for a magnet with perpendicular magnetic anisotropy (along $z$ direction) are shown.}
\label{figs2}
\end{figure*}
In this section, we provide a simulation framework that can be utilized to model the stochastic device physics of the spin-devices. The probabilistic switching characteristics of the ferromagnet can be analyzed by Landau-Lifshitz-Gilbert (LLG) equation with additional term to account for SOT generated by the heavy-metal (HM) underlayer \cite{slonczewski1989conductance},
\begin{eqnarray}
\frac{d\widehat{m}}{dt}=-\gamma (\widehat{m} \times H_{eff}) + \alpha (\widehat{m} \times \frac{d\widehat{m}}{dt})\nonumber + \frac{1}{q{N_{s}}}(\widehat{m} \times I_{s} \times \widehat{m})
\end{eqnarray}
where, $\widehat {\textbf {m}}$ is the unit vector of Free Layer (FL) magnetization, $\gamma= \frac {2 \mu _B \mu_0} {\hbar}$ is the gyromagnetic ratio for electron, $\alpha$ is Gilbert\textquoteright s damping ratio, $\textbf{H}_{eff}$ is the effective magnetic field including the shape anisotropy field, $N_s=\frac{M_{s}V}{\mu_B}$ is the number of spins in free layer of volume $V$ ($M_{s}$ is saturation magnetization and $\mu_{B}$ is Bohr magneton), and $\textbf{I}_{s}$ is the spin current generated by the HM underlayer. Thermal noise is included by an additional thermal field \cite{scholz2001micromagnetic}, $\textbf{H}_{thermal}=\sqrt{\frac{\alpha}{1+\alpha^{2}}\frac{2k_{B}T_{K}}{\gamma\mu_{0}M_{s}V\delta_{t}}}G_{0,1}$, where $G_{0,1}$ is a Gaussian distribution with zero mean and unit standard deviation, $k_{B}$ is Boltzmann constant, $T_{K}$ is the temperature and $\delta_{t}$ is the simulation time-step. Additional effects like considering field-like torque and Dzyalohinskii-Moriya Interaction (DMI) can be also included in the modelling framework \cite{emori2013current,kim2012layer}.

Fig. \ref{figs2} depicts two independent stochastic LLG simulations. The magnet was taken to be circular in shape with diameter $40nm$ and thickness $1.3nm$ with a barrier height of $31.44k_{B}T$. The magnet damping factor was taken to be 0.0122 \cite{pai2012spin}. A spin-Hall angle of 0.12 \cite{liu2012spin} was assumed for the Tantalum heavy metal layer of thickness $10nm$. The magnet was subjected to an in-plane magnetic field of strength $100Oe$ and a current pulse ($I$) of magnitude $900\mu A$ and duration $2ns$. As shown in the figure, the magnet stochastically relaxes to either of the two stable magnetization directions. Note that this section serves to outline the simulation framework that can be used to model the stochastic magnetization dynamics. We directly use the experimental probability switching characteristics obtained from Fig. \ref{fig1} for the Bayesian Network implementation discussed next.

\section*{Bayesian Network based on Stochastic MTJ}

\begin{figure*}[b]
\centering
\includegraphics[width=7.0in]{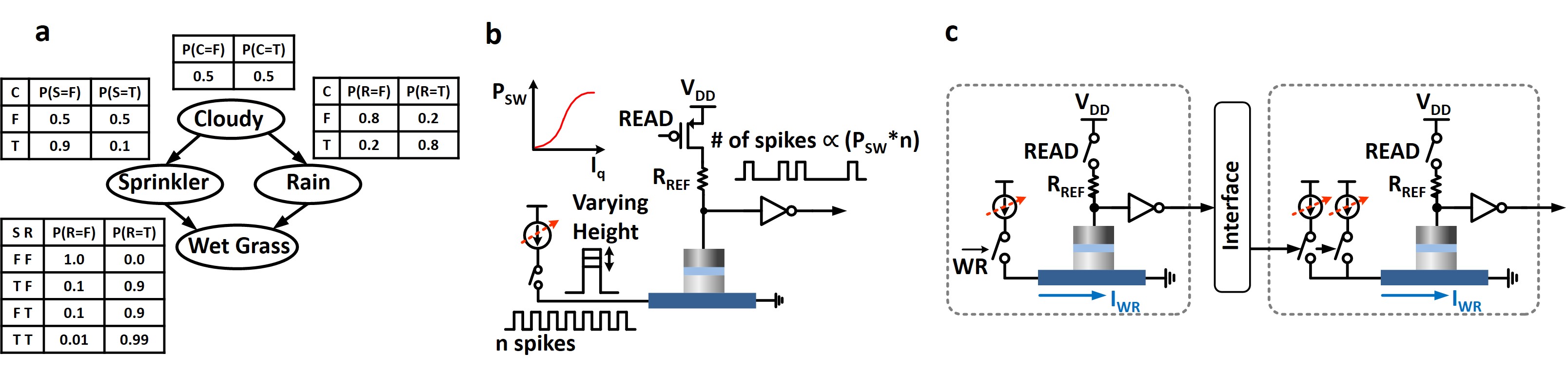}
\caption{(a) A simple Bayesian Network with 4 variables along with Conditional Probability Table (CPT) to show independence between variables \cite{kevin2001graphical}. (b) Generation of Poisson pulse train based on stochastic switching of nanomagnet with aid of CMOS peripherals. (c) Information transfer between neighboring variables through CMOS interface circuits.}
\label{fig4}
\end{figure*}

Bayesian Network (BN) is a graphical model to represent conditional independencies between each variable, where the nodes in a BN represent random variables and links represent direct dependencies among the variables \cite{jensen1996introduction,heckerman1995learning,nielsen2009bayesian}. 
A simple Bayesian Network is shown with four variables in Fig. \ref{fig4}(a) \cite{kevin2001graphical}. It illustrates dependencies among the variables -- whether it is cloudy (`C'), whether it is rainy (`R'), whether the sprinkler is on (`S'), and whether the grass is wet (`W'). Here we assume that each variable is binary (True or False, could be `1' or `0').
As shown in Fig. \ref{fig4}(a), the dependencies between variables are quantified using conditional probabilities (in Conditional Probability Table (CPT)) associated with each transition to a particular node from its parent nodes in the network. Due to its simple conditional independence statement expressed in graph, BN helps to reduce the number of variables required to compute a probabilistic inference. 
Based on the basic BN, the inference operation tries to estimate the probability of the hidden causes, on the given observed situation \cite{kevin2001graphical}. 
As an example, let us assume that we observe wet grass and attempt to estimate the cause. There are possibly two hidden causes, either the Sprinkler is on or it is raining. Here we can use Bayes' rule (shown below) to calculate posterior probability of each cause,
\begin{eqnarray}
P\left ( A|B \right )=\frac{P\left ( B|A \right )P\left ( A \right )}{P\left ( B \right )}
\end{eqnarray}
where A are the hidden causes and B is the observed evidence. 
Implementation of a Bayesian inference network on conventional general-purpose computers is inefficient in terms of area and energy consumption since a large number of complex floating point calculations need to be performed to compute the probability of occurrence of a particular variable since multiple causal variables are involved in the network. 
Hence, various approaches for BN hardware implementation have been proposed based on synthesizable hardware such as Field Programmable Gate Arrays \cite{lin2010high,zermani2015fpga}, fully digital system with stochastic digital circuit \cite{mansinghka2008stochastic,thakur2016bayesian}, analog based probabilistic hardware for inference \cite{mroszczyk2014accuracy,weijia2007pcmos}, and mixed-signal approach (Muller C-Elements) \cite{friedman2016bayesian}. 
However, multiple transistors are required to implement the functionality of a single stochastic element, thereby leading to an inefficient design. In this work, we demonstrate the manner in which the stochastic switching behavior of ferromagnet-heavy metal structures can directly mimic the core behavior of such controllable random number generators.

\subsection*{Poisson spike generation based on stochastic switching of nanomagnet}

Instead of using conventional floating point calculations to estimate the probability of a certain inference process, pulse based computation \cite{murray1988pulse} can be exploited. Here the probability of inference operation can be estimated by counting the number of pulses from each variable over a large enough time window. The main idea behind this approach is that the variable of the BN is represented by a Poisson pulse train generator that converts the probability information into the frequency of the output pulses. 
Based on the controllable stochastic switching of the nano-sized magnet with thermal noise, the Poisson spikes can be generated with the aid of simple CMOS peripherals as shown in Fig. \ref{fig4}(b). Here the device is interfaced with a reference resistor to generate a Poisson spike/pulse train where the number of spikes in a large enough time window encode information about the frequency and magnitude of the incoming spike train. The operation can be explained by considering a write/ read/ reset cycle. Let us consider the case for a single stochastic element in Fig. \ref{fig4}(c) that receives input from another stochastic element. Note that each element represents a particular node in the BN. During the write phase, the causal stochastic element which is transmitting spikes from the previous stage inverter generates a write current $I_{WR}$ through the heavy metal. The magnitude of the write current is determined by the corresponding current source which is tuned according to the CPT. The current source is activated with a frequency equivalent to the frequency of the incoming spike train. After the write cycle, the read cycle is used to determine whether the device has switched and is reset to the initial state in case of a switching event. Hence, the frequency of the output train is directly proportional to the switching probability (determined by the magnitude of the current source which is tuned depending on the corresponding entry in the CPT) and the pulse train frequency from the causal element. The pulse train frequency from the causal stochastic element encodes the probability of occurrence of the corresponding event while the current source being driven by that element encodes the conditional probability of occurrence of the receiving element. 

\subsection*{Information transfer through the network}

\begin{figure*}
\centering
\includegraphics[width=7.0in]{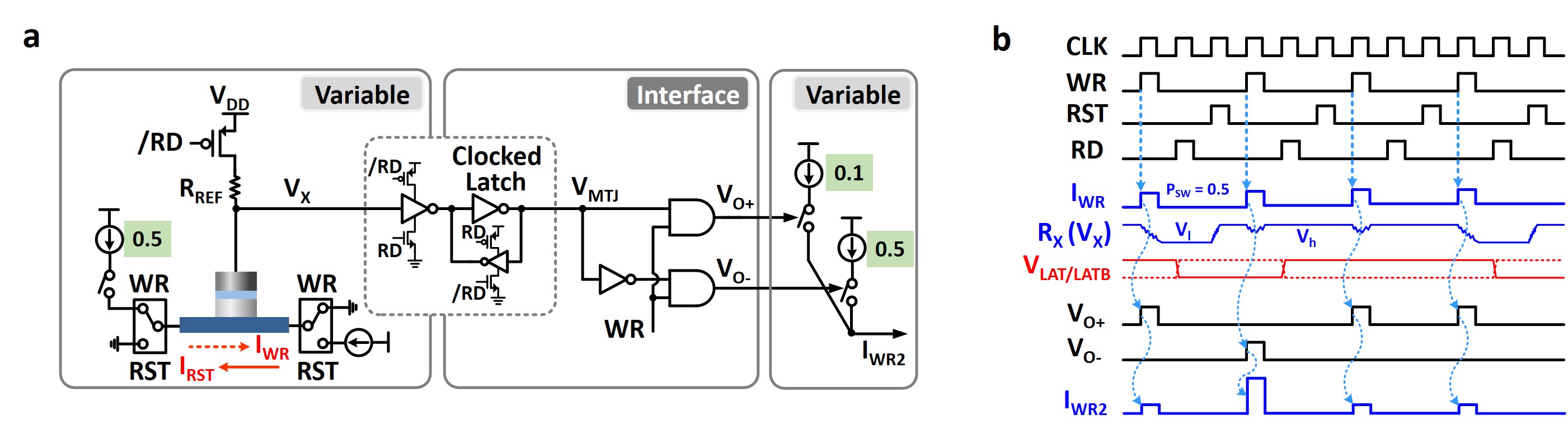}
\caption{(a) Detailed implementation view of BN variable and interconnection between neighbors. (b) Timing diagram used to perform probabilistic estimation based on the proposed BN implementation.}
\label{fig5}
\end{figure*}

The detailed implementation of the core components (variable and edge) proposed for the BN is shown in Fig. \ref{fig5}(a). The MTJ device along with two current sources and few other circuit elements constitute each variable of the BN. Note, here we depict two current sources for each of Write (WR) and Reset (RST) operations. Two switches at either sides of the nanomagnet control the direction of current flow during each operation. 
The inverter at the output node for level conversion is combined with a latch and forms a Clocked Latch (C-LAT). This Clocked Latch operates in sync with the Read (RD) signal and can amplify small voltage changes at the $V_x$ node. In addition, this unit stores the value from the result of stochastic switching. 
This provides two important benefits: 1) It prevents an interaction from the changes in voltage during the Read operation to the rest of the network. Without this intermediate storing stage, unwanted analog voltage fluctuation during the Read operation can be amplified and be transmitted to the next stage. 2) Also, it makes the system operate in a pipelined manner, and hence contributes to make the system synchronous. The variables in BN except for the root variable can only process information on receiving the data from the parent node, which makes the system asynchronous with low throughput. By adopting clocking with proper storing element, the system become synchronous and can have high throughput. 
The stored value in C-LAT represents result of current stochastic switching. When the output level of C-LAT ($V_{MTJ}$) is combined with the Write command, the output pulse from each node is generated. Note that, depending on the level of $V_{MTJ}$, either of $V_{o+}$ or $V_{o-}$ can generate a pulse at a time. In case the level of $V_{MTJ}$ is logic `1', then only $V_{o+}$ node can generate a pulse when the Write pulse is presented to the AND gates. On receiving this pulse, corresponding current pulse at the next stage is generated.

Abovementioned operations are illustrated with a timing diagram in Fig. \ref{fig5}(b). Based on global clock signal (CLK), three main pulses are generated for Write, Reset, and Read operations (WR, RST, and RD). These three signals are broadcasted through global signaling and used by all the variables and interfaces. 
Let us assume that the initial magnetization direction is parallel configuration (lower resistance than $R_{REF}$), hence $V_{x}$ node is at a high voltage level ($V_h$). Since the switching probability of the initial node is set to be 0.5, the magnet flips its state with $50\%$ chance. The $V_x$ node reflects such a switching frequency and can have two possible states ($V_h$ and $V_l$). This small voltage difference is amplified and stored in the Clocked Latch ($V_{LAT}/V_{LATB}$) on receiving the Read command pulse. 
The stored value - stochastic switching result - is now converted into a pulse when this value is combined with the global Write command. Depending on the switching results, either $V_{o+}$ or $V_{o-}$ transmits a pulse to the next variable, and thereby can generate a current pulse with desired amplitude as we specified through the CPT.

\subsection*{Implementation of BN and inference operation}

\begin{figure*}
\centering
\includegraphics[width=6.0in]{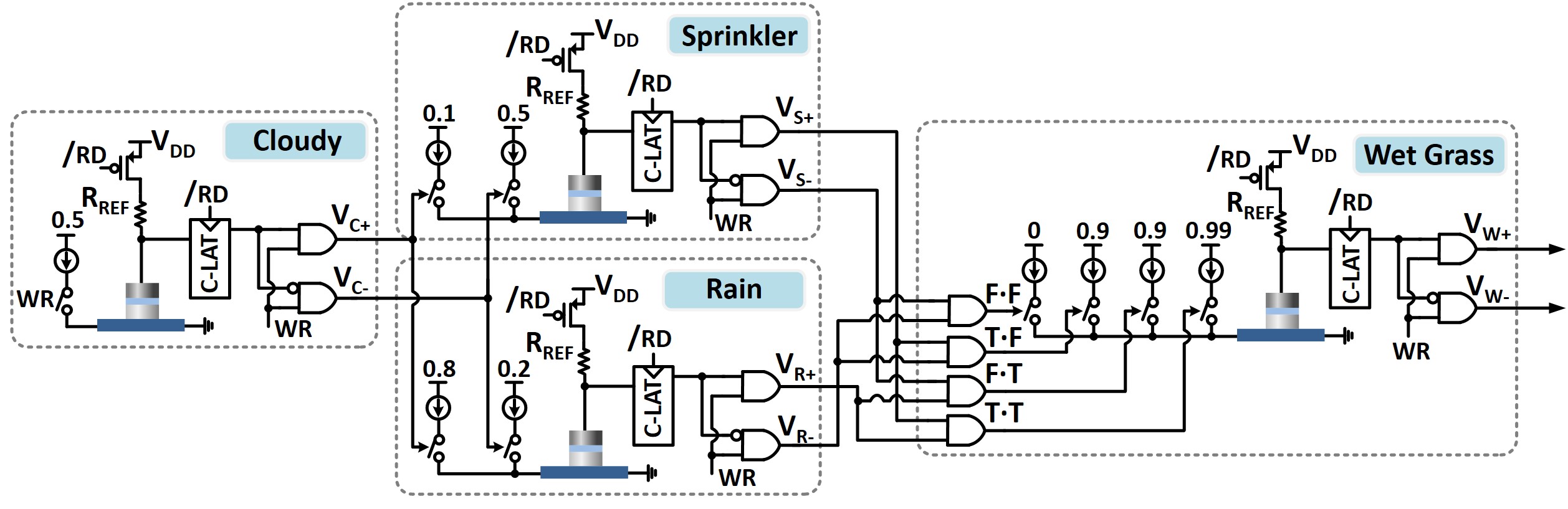}
\caption{Implementation of the BN with 4 variables based on the proposed device-circuit configuration.}
\label{fig6}
\end{figure*}

Based on the implementation of the variable and edge in the previous section, we discuss the implementation of the entire network (in Fig. \ref{fig4}(a)). Fig. \ref{fig6} shows a complete view of the BN implementation using the proposed device-circuit configuration. 
Each variable is made up of a single stochastic device with the required peripherals. The next state of each variable is determined by controllable stochastic switching which contributes to the generation of a Poisson spike train. The switching frequency information is transferred to the next variable as a pulse through the aforementioned interface circuitries. Note that, four AND gates are adopted at the beginning of the last variable to generate a multiplication output between two Poisson spike trains.
Based on the complete network in Fig. \ref{fig6}, we can infer the probability of each variable, i.e. P(Sprinkler), P(Rain), and P(Wet), which is difficult to get directly from the CPT. This could be accomplished just by counting the number of output pulses corresponding to each variable for a long enough time duration. For instance, if 28 pulses are counted at the $V_{s+}$ output of `Sprinkler' node over 100 write cycles, this means that the probability of `Sprinkler is on' is estimated to be $28\%$. 

\begin{figure*}
\centering
\includegraphics[width=7.0in]{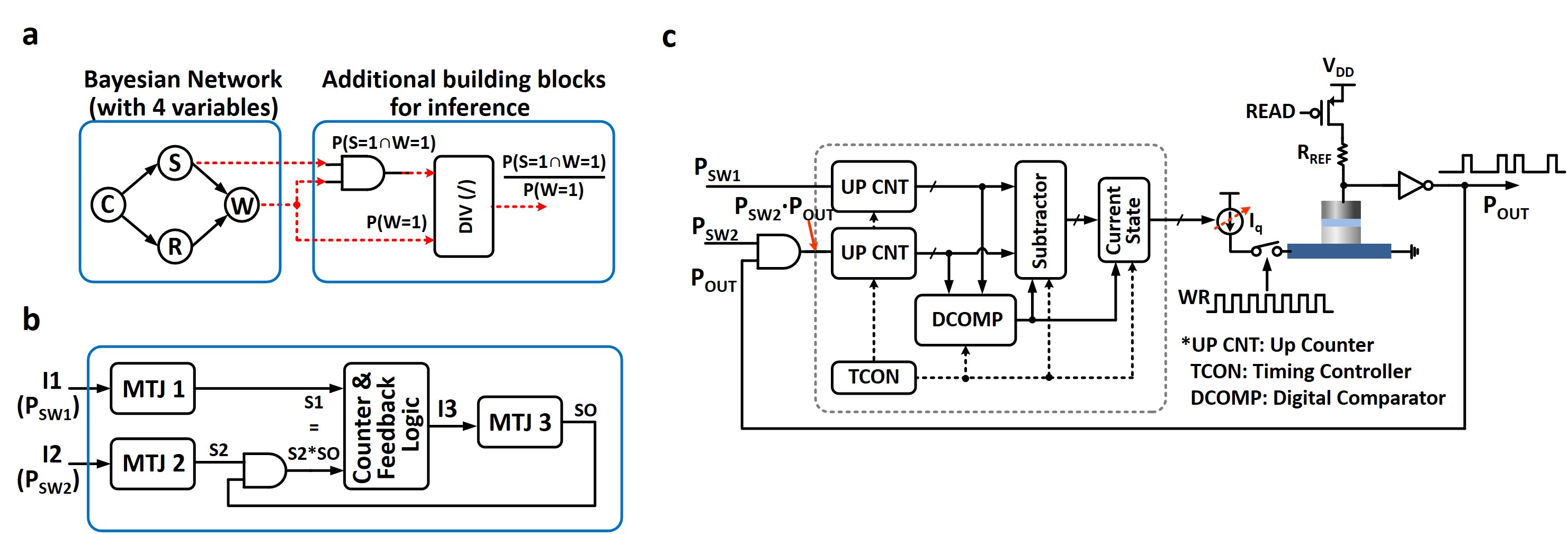}
\caption{(a) The inference operation is performed through the division and multiplication operation between two Poisson pulses based on the complete network in Fig. \ref{fig6}. (b) Division operation by matching the rate of spiking between two nodes \cite{thakur2016bayesian}. (c) Practical implementation of the divider based on stochastic MTJ and CMOS interface circuits.}
\label{fig7}
\end{figure*}

Moreover, estimation of more complex inference is also possible by introducing additional arithmetic building blocks such as division and multiplication between two Poisson pulses (Fig. \ref{fig7}(a)). 
Let us consider the same question that was mentioned in the earlier section, i.e. the probability of `Sprinkler is on' in a given situation of `Wet grass is true'.
By the definition of conditional probability, this could be written as $P(S=1|W=1) = P(S=1\cap W=1)/P(W=1)$. Here the intersection function between the two probabilities can be interpreted as multiplication between two Poisson spike trains which could be implemented by an AND gate. The only remaining function to be implemented is the division operation.
Recently, the method of performing division operation between two Poisson spike trains by matching the rate of spiking of two internal signals has been proposed in Ref.\cite{thakur2016bayesian}. Fig. \ref{fig7}(b) depicts the concept of the division operation between two Poisson spikes and practical implementation of the divider is shown in Fig. \ref{fig7}(c). 
Let us first describe the division operation using Fig. \ref{fig7}(b). There are two input pulse trains with different spiking rate to the division unit (with current inputs $I1$ and $I2$ to the proposed device-circuit configuration, named as MTJ1, MTJ2). Here we denote spiking rate of upper and lower pulse trains as `S1' and `S2' respectively. Likewise, the rate of spiking from the output node is denoted as `SO'. Note, the current input pulse to the output MTJ device (MTJ3) is generated from additional CMOS logic gates (named as `Counter \& Feedback Logic'). 
By multiplying the two pulses, `S2' and `SO', using AND gate, the resultant spiking rate becomes `S2*SO'. Here the main idea is that once the `Counter \& Feedback Logic' controls the spiking rate of MTJ3 device to match the spiking rate of `S1' and `S2*SO' to be the same, then the rate of output spiking becomes $SO =S1/S2$. 
For this functionality, the Counter logic counts the number of pulses from each of `S1' and `S2' over a predetermined time duration. If the comparison results show `S1' is more than `S2', then the Feedback logic raises the spiking rate of output `SO' so that `S2*SO' rate also increases. By following this simple feedback rule, the division operation can be achieved. Fig. \ref{fig7}(c) shows the implementation of the divider unit based on this approach. The device-circuit configuration to generate an output Poisson spike train is identical to the one we used to implement the variable in BN. The only difference is the rate of spiking (i.e. amplitude of input current pulse), is controlled by the `Counter \& Feedback Logic' inside the dotted box. The `Counter \& Feedback Logic' consists of two digital up-counters to count the number of spikes from two inputs, digital comparator, subtractor, and timing controller (TCON).

\section*{Results}

\begin{figure*}
\centering
\includegraphics[width=7.0in]{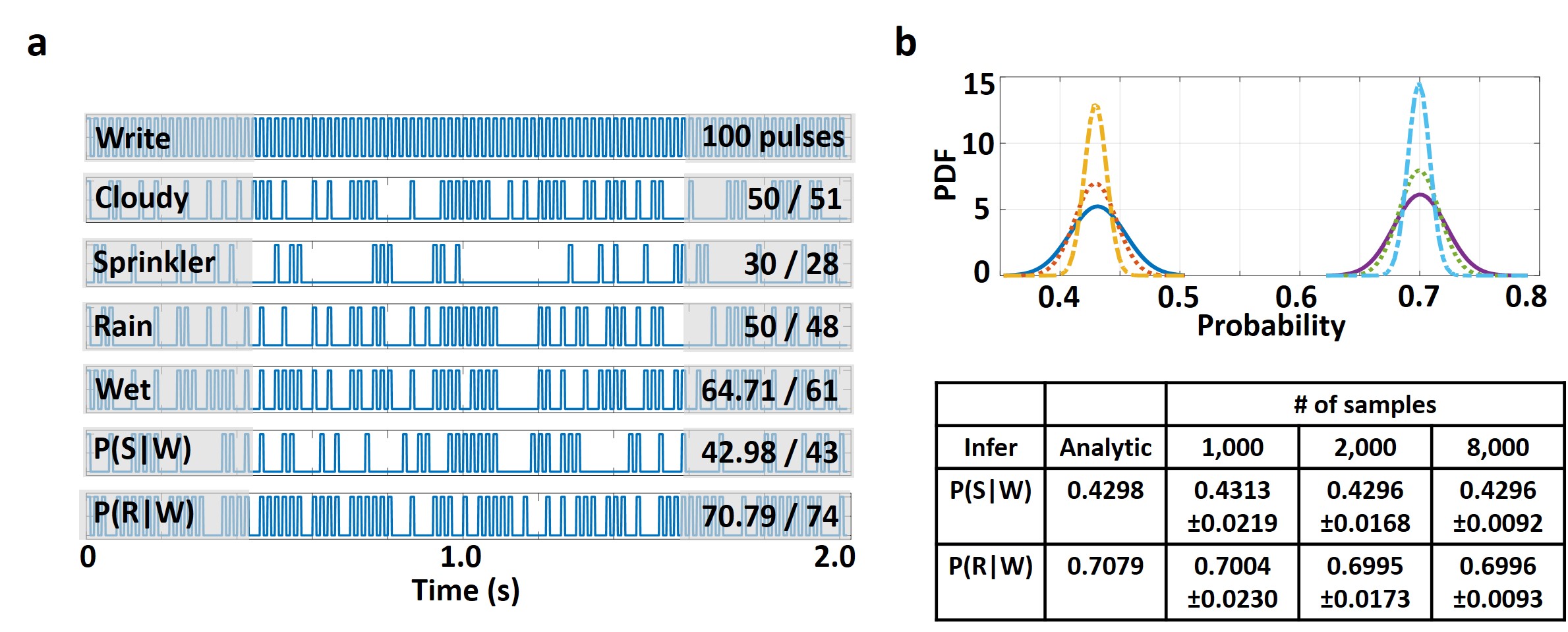}
\caption{(a) Timing waveform for the calculation of probability of occurrence of various events. The average number of spikes produced in each case over 100 sample points closely resemble the actual analytical values. Data represented in the format A/B denote that A is the analytical solution while B is the computed probability from the 100 sample points of each output. (b) The probability distribution function (PDF) of the two conditional events $(S|W)$ and $(R|W)$ approach the mean value as the inference samples increase.}
\label{fig8}
\end{figure*}

The functionality of the BN was verified by performing a simulation of the entire network based on the probability switching characteristics depicted in Fig. \ref{fig1}(b). Fig. \ref{fig8}(a) represents the timing waveforms and average number of spikes produced for various events which approximates the actual analytical solution to a reasonable degree of precision. Fig. \ref{fig8}(b) depicts that the probability distribution function (PDF) of the two conditional events $(S|W)$ and $(R|W)$ approach the actual analytical solution as the number of samples increases. 

It is worth noting here that Bayesian inference and in general, the class of unconventional computing platforms that can be enabled by such stochastic magnetic elements, are inherently resilient to variations and noise in the computational units. For instance, we perform an analysis of the network performance in presence of random Gaussian transient noise added to the input charge current provided by the CMOS current source through the heavy metal layer. 
Note that $\sim60\mu A$ ($0.48-0.54 mA$) of current range is being exploited from the switching probability characteristics of the magnetic stack (Fig. \ref{fig1}b) for proper functioning of the Bayesian network. We utilize a 6-bit DAC to provide the input current through the heavy metal, thereby providing $\sim1\mu A$ of current resolution (LSB). 
Table \ref{tab:example} depicts the mean and standard deviation of the output $P(R|W)$ of the network for $1,000$ samples with varying amplitudes of the Gaussian noise (from $1LSB\sim1\mu A$ to $3LSB\sim3\mu A$). As expected, the mean value remains close to the case without noise. Although the variance of the PDF distribution increases by a small amount, it can be reduced by increasing the number of sample points used for inference (Fig. \ref{fig4}b).

\begin{table}[ht]
\centering
\begin{tabular}{|c|c|c|}
\hline
Noise amplitude & Mean of $P(R|W)$ & STD ($\sigma$) of $P(R|W)$ \\
\hline
0 (No noise) & 0.7048 & 0.0222 \\
\hline
1 LSB & 0.7046 & 0.0249 \\
\hline
1 LSB & 0.7046 & 0.0339 \\
\hline
1 LSB & 0.7031 & 0.0545 \\
\hline
\end{tabular}
\caption{\label{tab:example}Estimated probability of the inference for 1,000 output samples with Gaussian noise.}
\end{table}


\section*{Discussion}

Prior proposals have considered experimental demonstration of spin-orbit torque driven magnetic heterostructure switching for unconventional computing platforms like associative memory operations \cite{borders2016analogue}. However, the proposals typically exploit analog and deterministic Hall-bar switching (the resistance of the Hall-bar varies in an analog fashion depending on the magnitude of the input current) as the core computing element. 
In contrast, we are exploiting binary and stochastic Hall-bar switching for our proposal. Such probabilistic computation is expected to replace deterministic computing platforms (based on such post-CMOS technologies) since at highly scaled dimensions such devices are not expected to exhibit multi-bit analog resolution. In contrast, stochasticity will become increasingly predominant. Therefore, exploring probabilistic computing platforms based on stochastic device switching that encode information in time (through probabilistic update of binary elements) rather than space (through deterministic analog computing elements) will become important. This work can potentially stimulate efforts at developing stochastic computing platforms that embrace the underlying stochasticity of highly-scaled nanomagnets. 

In conclusion, in this article we provided proof-of-concept experiments demonstrating probabilistic spin-orbit torque induced magnetization reversal. Such stochastic devices can provide a direct mapping to the computing elements of Bayesian inference, Deep Belief Networks and probabilistic neuromorphic applications.


\section*{Methods}
\subsection*{Sample Fabrication}
The device was fabricated by utilizing two consecutive steps of e-beam lithography. Silicon with $\sim 2000A$ thick thermal oxide was used as the substrate. The first step resulted in the development of the Hall cross structure while the second step was used for the fabrication of the contact pads. One layer of polymethyl methacrylate (PMMA) e-beam resist was coated onto the surface of a clean silicon wafer surface. Then e-beam lithography was implemented to define the Hall bar structure for the device fabrication. After development, the Hall bar was deposited by magnetic sputtering. The stack structure developed was Ta (10 nm)/CoFeB (1.3 nm)/MgO (1.5 nm)/Ta (5 nm) (from bottom to top). A piece of cleaned bare wafer was also deposited in-situ so that the thin film deposited on it could be considered with same magnetic property as that on patterned chip. This piece was measured on vibrating sample magnetometer (VSM) for magnetic properties such as magnetic saturation, coercivity, etc. After lift-off for the patterned chip, a similar fabrication process of contact pads was followed. The major difference was that the contacts were made of Ta (10 nm)/Au (120 nm) by e-beam evaporation. After lift-off, the final Hall-cross device structure can be seen in Fig. \ref{fig1}.

For each of the samples investigated for this probability study, magnetic properties were characterized by VSM on the in-situ fabricated bare wafer. The saturation magnetization of the samples was measured to be $\approx 581.36$ $e.m.u. cm^{-3}$ and the coercivity was $\approx 5.07$ $Oe$. All samples showed perpendicular magnetic anisotropy (PMA). Current induced magnet switching and probability measurement were performed on a probe station equipped with an in-plane magnetic field. 

\section*{Acknowledgements}

The authors would like to thank Prof. Zhihong Chen of Purdue University for advice regarding the device fabrication and measurements. The work was supported in part by, Center for Spintronic Materials, Interfaces, and Novel Architectures (C-SPIN), a MARCO and DARPA sponsored StarNet center, by the Semiconductor Research Corporation, the National Science Foundation, Intel Corporation and by the US DoD Vannevar Bush Faculty Fellowship.

\section*{Author contributions statement}

S. Chen fabricated the samples and performed the measurements. Y. Shim performed the simulations for the Bayesian inference framework based on the probabilistic switching characteristics obtained from the samples. All authors assisted in discussing the results and writing the manuscript.

\section*{Additional information}

\textbf{Competing financial interests}: The authors declare no competing financial interests. 





\end{document}